\documentclass{IEEEtran}
\usepackage{cite}
\usepackage{amsmath,amssymb,amsfonts}
\usepackage{algorithmic}
\usepackage{graphicx}
\usepackage{textcomp}
\usepackage{tikz}
\usepackage{pgfplots}
\pgfplotsset{compat=1.18}
\usetikzlibrary{arrows.meta, decorations.pathreplacing}
\usepackage[caption=false,font=footnotesize]{subfig}
\begin{document}
\title{Radial-Angular Resolution Framework for Near-Field Communications}
\author{{\IEEEauthorblockN{Viviana Centritto, Sotiris Droulias and Angeliki Alexiou,~\IEEEmembership{Member,~IEEE}} }
\thanks{The authors are with the Department of Digital Systems, University of Piraeus, Piraeus 18534, Greece (Corresponding author: Viviana Centritto, \mbox{e-mail:vcentritto@unipi.gr).}}}

\maketitle

\begin{abstract}
Near-field beamfocusing confines radiated power into a localized three-dimensional focal spot, making the analytical characterization of radial and angular resolution critical for sixth-generation (6G) multi-user systems, yet existing models treat the focal resolution as a fixed geometric outcome rather than a tunable design variable. Based on the Angular Representation (ASR) approach, this paper presents a resolution framework to design contiguous focal regions across both the radial and angular domain for steered multi-user systems. By extending the definition of a tunable inter-beam overlap $\rho$, we derive closed-form expressions for the radial and angular focal boundaries and extents, alongside the steering-dependent minimum focal distance $d_{0,\text{min}}$ and maximum steering angle $\theta_{\text{max}}$. Using these boundaries, iterative synthesis algorithms are developed to construct uniform and non-uniform radial and angular beam arrangements. Finally, by analyzing both isolated and joint radial-angular configurations, we quantify the fundamental trade-offs between $\rho$, achievable user capacity, inter-user interference, and system sum-rate across varying steering angles.

\end{abstract}


\section{Introduction}
\label{sec:introduction}
 As sixth-generation (6G) networks migrate toward sub-terahertz (sub-THz) carrier frequencies and deploy extremely large antenna arrays (ELAAs) spanning hundreds or thousands of wavelengths, the Fraunhofer distance $2D^2/\lambda$ expands to tens or even hundreds of meters \cite{cui2022near}–\cite{selvan2017fraunhofer}. Consequently, a substantial portion of communication links, involving users, sensors and environmental scatters, will operate within the radiating near-field region. Unlike the far-field regime, where the planar wavefront assumption limits spatial multiplexing to the angular domain, near-field propagation is characterized by spherical wavefronts. This phase curvature unlocks the radial domain as an additional spatial degree of freedom, enabling beamfocusing. Rather than steering the beam along a semi-infinite angular direction, this capability allows the radiated power to be confined into localized, three-dimensional focal spots \cite{zhang20236g, zhang2022beam}. By exploiting this 3D spatial confinement, beamfocusing serves as a critical enabler for range-domain multiple access, high-resolution sensing, and precise localization \cite{cong2024near, ouyang2024primer}

Nevertheless, the efficacy of beamfocusing in emerging multiple-access techniques depends on how these focal regions are characterized in terms of spatial resolution. Separating co-directional users requires stringent confinement in the radial domain to mitigate longitudinal interference, while isolating users across different directions relies on precise angular resolution. This challenge is exacerbated when beams are steered away from the aperture normal; at oblique angles, the effective aperture diminishes, leading to the distortion and elongation of the focal spot. Therefore, dimensioning multi-user topologies need robust analytical models capable of characterizing the spatial boundaries of contiguous focal regions under steered conditions. Such models are essential for deriving design insights for next-generation network architectures.

\subsection{Prior Works}
Near-field beamfocusing concentrates the radiated power at a finite depth, allowing users along one direction to be separated by distance \cite{zhang20236g,zhang2022beam}. The resolution of the resulting focal regions can be described analytically using the Angular Spectrum Representation approach (ASR). This approach was first applied to Reconfigurable Intelligent Surfaces (RIS) to model beam steering and focusing in \cite{droulias2022reconfigurable}, and it was later extended to oblique incidence and reflection angles in \cite{droulias2024near}, providing closed-form expression for the spatial power density alongside  key performance metrics, such as the Full-Width-Half-Maximum (FWHM), which quantifies the spatial extent of the focal spot. In \cite{centritto2026analytical}, this metric was analytically expanded to facilitate the design of contiguous focal regions, featuring non-uniform or tailored resolution, along the radial domain via a tunable inter-beam overlap parameter $\rho$. While the framework provides guidelines for dimensioning beamfocusing systems, it remains confined to the propagation axis, assuming beams are focused solely along the aperture normal.

Alternatively, other works rely on the Fresnel approximation of the Huygens-Fresnel integral. For instance, the authors in \cite{bjornson2021primer} and \cite{kosasih2024finite} characterize the radial extent of the focal spot by defining the 3 dB beam depth (BD) metric, which expands with the focal distance and depends on both the array geometry and the observation angle. Subsequent studies refine this metric by establishing the Effective Beamfocusing Rayleigh Distance (ERBD), an angle-dependent boundary that determines the maximum range for effective beamfocusing across linear, rectangular and circular arrays \cite{hussain2026analyzing} \cite{hussain2026uniform}. By revisiting the Fresnel-Kirchhoff diffraction integral, the authors in \cite{manohar2024resolution} derive closed-form expressions for transverse and axial 3 dB resolutions of near-field focused arrays as a function of array size, focal distance and operating frequency. Furthermore, recent studies in sensing leverage the BD metric and the ambiguity function to compare range resolution across array geometries and SIMO/MIMO configurations \cite{wachowiak2025sizing}, \cite{wachowiak2026analysis}. 

A separate body of literature investigates how multiple focused beams can coexist to serve multiple users, mostly through their correlation rather than their spatial extent. Within this domain, an initial line of research characterizes the resolution of these focused beams. In \cite{wu2022multiple}, the authors demonstrate that focused beam correlation vanishes asymptotically with increasing array size, motivating the development of Location Division Multiple Access (LDMA). Conversely, \cite{rao2024general} analytically confirms that spatial resolution remains imperfect at distances comparable to the Rayleigh range, laying the foundation for Non-Orthogonal Multiple Access (NOMA)-schemes that exploit shared beams. A second line instead designs beamsets to establish non-interfering communication modes with minimal mutual correlation. By applying the Rayleigh criterion to align beam peaks with adjacent nulls, \cite{decarli2021communication} defines such modes for linear apertures; however, this inherently fixes the beam spacing to the focal spot width, leaving the inter-beam overlap threshold as a passive outcome rather than a tunable parameter. Additionally, recent works by \cite{droulias2025orthogonal, droulias2025orthogonalcodebook} establish that finite-size transmitters can achieve strictly zero correlation via circular arrays or cosine beams. Although these methods generate orthogonal sets across both range an angle, the maximum number and placement of allowable of these beams is constrained by null-spacing conditions.

Taking together, existing literature predominantly treats the spatial extent of focal regions as a fixed geometric consequence dictated by the array shape and focal distance, rather than as a tunable design variable. In multi-user scenarios, inter-beam spacing is often constrained by orthogonality requirements, null-spacing conditions, or an overlap-based dimensioning restricted to the broadside axis. For steered beams, although the observation angle geometrically affects the 3 dB BD, current models do not jointly characterize the radial and angular boundaries of the focal regions based on a prescribed overlap. Consequently, a comprehensive framework that utilizes controllable inter-beam spacing to define these boundaries across both the radial and angular domain remains absent.

\subsection{Our Contributions}
By leveraging the ASR modeling approach, this work introduces a framework for steered multi-user systems. Specifically, we extend the inter-beam overlap parameter $\rho$ defined in \cite{centritto2026analytical} to map directly to the radial and angular boundaries of each focal region, ultimately translating these controllable spatial resolutions into maximum user capacity and inter-user interference levels. The main contributions of our work are:
\begin{itemize}
    \item We analytically derive the radial boundaries $r_\text{start}$ and $r_\text{end}$ of a focal region along the steering direction $\theta_r$. This defines the radial focal extent $\Delta r_\rho$ and introduces a tunable inter-beam overlap parameter $\rho_\text{radial}$, providing analytically control over the resolution of contiguous focal regions in the radial domain.
    \item We extend the framework to the angular domain by evaluating the power density along an arc of constant radius $r_\text{peak}$. We derive closed-form expressions for the angular boundaries $\theta_\text{start}$ and $\theta_\text{end}$, as well as the angular extent $\Delta \theta_\rho$, enabling equivalent overlap control for contiguous beams across varies angles through $\rho_\text{angular}$
    \item Leveraging these closed-form radial and angular boundaries solutions, we develop a numerical synthesis methodology to contiguously chain focal regions that intersect at a tunable inter-beam parameter $\rho$.
    \item We analytically establish the steering-dependent spatial boundaries. By generalizing the paraxial approximation in \cite{centritto2026analytical} to oblique angles, we derive $d_{0,\text{min}}$. Furthermore, we derive the maximum steering angle $\theta_\text{limit}$ for a specific focal distance in the angular domain.
    \item We perform a system-level assessment to evaluate per-user and sum-rate spectral performance under the proposed framework. By analyzing both isolated and joint radial-angular configurations, we quantify the trade-offs between the inter-beam overlap parameter $\rho$, user capacity, individual per-user spectral efficiency and system sum-rate.
\end{itemize} 

\begin{equation} \label{eq:ASR}
    \mathbf{E}_r (x,y,z) = \iint_{-\infty}^{-\infty} \hat{\mathbf{E}}^{UPA}(k_x,k_y) e^{j(k_xx+k_yy + k_zz)} dk_xdk_y
\end{equation}
\begin{figure}[t!]
    \centering
    \includegraphics[width=1.0\columnwidth]{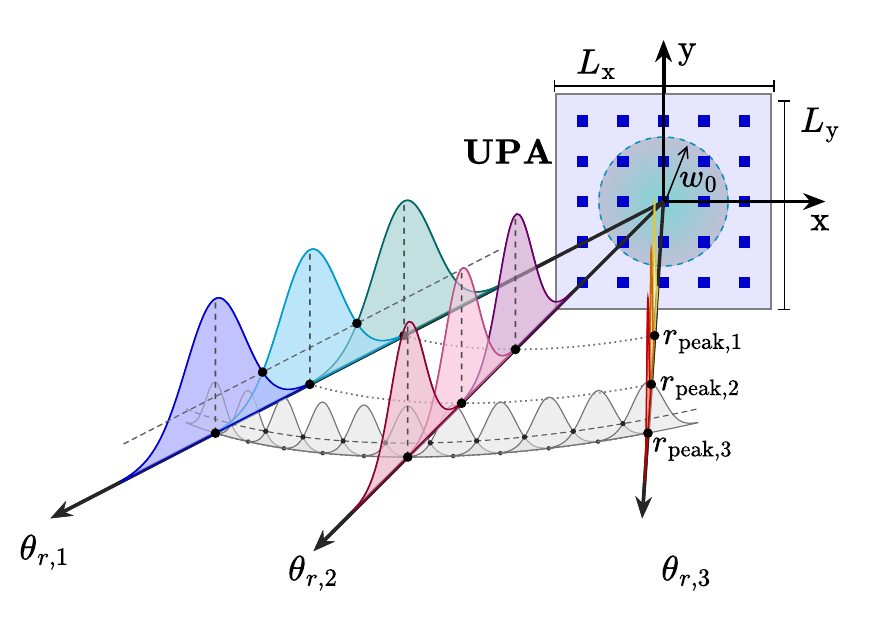}
    \caption{System model of an $L_x \times L_y$ UPA in the $xy$-plane with a beam radius $w_0$, showing contiguous focal regions synthesized along steering directions ($\theta_{r,1}, \theta_{r,2}, \theta_{r,3}$) and constant-radius arcs ($r_{\text{peak},1}, r_{\text{peak},2}, r_{\text{peak},3}$).}
    \label{fig:system_model}
\end{figure}

\section{Analytical framework and system model}
\label{sec:analytical_framework}
Let us consider a Uniform Planar Array (UPA) centered at the origin of the $xy$- plane, consisting of $L$ radiating elements, with $L_x$ and $L_y$ elements along the $x$- and $y$- directions, respectively. As illustrated in Fig. \ref{fig:system_model}, an incident wave with wavevector $\mathbf{k_i}$ impinges on the array surface at an angle of incidence $\theta_i$. The incident beam is assumed to be $y$-polarized with its footprint modeled by a Gaussian distribution. The incident field $\mathbf{E_i}$ is characterized by its magnitude $E_0$, free-space wavenumber $k =\frac{2\pi}{\lambda}$ (where $\lambda$ is the wavelength), and the beam radius $w_0$
\begin{equation}
    \mathbf{E}_i(x,y) = E_0 e^{\left( -\frac{(x\cos{\theta_i})^2 + y^2}{w_0^2} \right)} e^{(-jk\sin{\theta_i}x)}\mathbf{\hat{y}},
\end{equation}
while the footprint is circular under normal incidence ($\theta_i = 0^\circ$), it acquires an elliptical shape under oblique angles ($\theta_i \neq 0^\circ$). To realize beamfocusing, the UPA shapes the incident wave into spherical wavefronts that converge at a desired near-field location characterized by a focal distance $d_0$ and a steering angle $\theta_r$. The field distribution at the aperture of the UPA ($z=0$) can be expressed as
\begin{equation}
    \mathbf{E}^{UPA}(x,y) = \mathbf{E}_i(x,y) e^{j\phi(x,y)},
\end{equation}
where $\phi(x,y)$ represents the phase profile applied by the array to steer and focus the beam, defined as 
\begin{equation}
\resizebox{1\columnwidth}{!}{$\displaystyle
    \phi(x,y) = -k\sin{\theta_i}x + kd_0\cos{\theta_r} \times \sqrt{1+ \left( \frac{x-d_0\sin{\theta_r}}{d_0 \cos{\theta_r}}\right)^2 + \left(\frac{y}{d_0 \cos{\theta_r}}\right)^2}
$}
\end{equation}

To model the free-space propagation of the beam along the wavevector $\mathbf{k_r}$ within the $xz$- plane, we employ the ASR technique. This method facilitates the calculation of the field distribution $\mathbf{E}_r(x,y,z)$ at any observation plane ($z>0$) by determining the Fourier transform of the field $\mathbf{E}^{UPA}(x,y)$ (sources)
\begin{equation}
    \hat{\mathbf{E}}^{UPA}(k_x,k_y) = \iint_{-\infty}^{-\infty} \mathbf{E}^{UPA}(x,y) e^{-j(k_xx+k_yy)} dxdy,
\end{equation}
where $k_x$ and $k_y$ represent the spatial transversal frequencies. As detailed in \cite{droulias2024near}, the radiated field must satisfy the Helmholtz equation $(\nabla^2 + k^2) \mathbf{E}_r = 0$. Substituting the Fourier representation of $\mathbf{E}_r$ into this equation reveals that the field propagates according to $\hat{\mathbf{E}}_r(k_x,k_y;z) = \hat{\mathbf{E}}^{UPA}(k_x,k_y) e^{jk_zz}$, where $k_z = \sqrt{k^2 - k_x^2 - k_y^2}$ represents the longitudinal spatial frequency. Performing the inverse Fourier transform yields the angular spectrum representation
\begin{figure}[!t]
    \centering
    \subfloat[\label{fig:xz_three_beams}]{%
        \includegraphics[width=0.9\columnwidth]{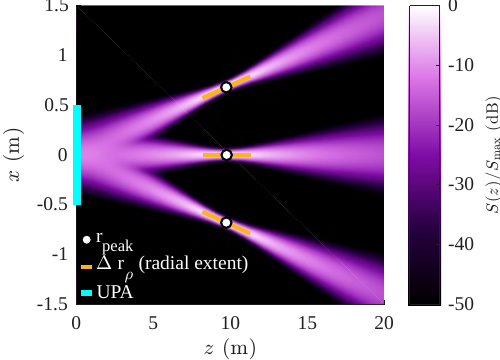}%
    }
    \\ \vspace{1em} 
    \subfloat[\label{fig:single_beam_radial}]{%
        \includegraphics[width=0.5\columnwidth]{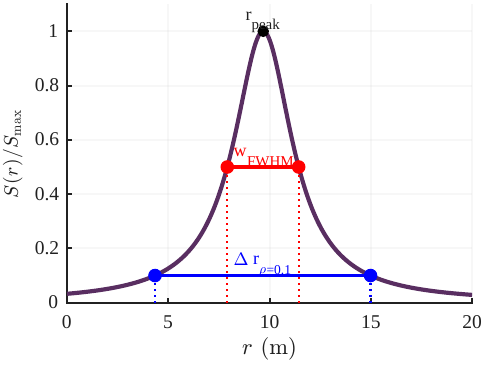}%
    }
    \hfill
    \subfloat[\label{fig:single_beam_angular}]{%
        \includegraphics[width=0.5\columnwidth]{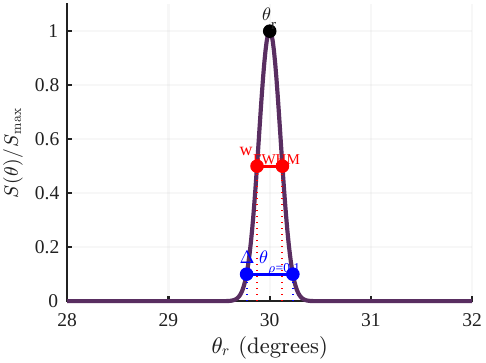}%
    }
    \caption{Radial-angular characterization of near-field focused beams, illustrating the normalized power destiny distribution the steering-direction. (a) Cross-section for three steered beams in the $xz$-plane. (b) Profile along the steering axis $\theta_r$. (c) Profile along a constant-arc $r = r_{\text{peak}}$}.
    \label{fig:beam_cross_section}
\end{figure}
Finding a direct analytical solution to (\ref{eq:ASR}) is highly complex due to the square root terms present in the exponents of both $k_z$ and the aperture phase profile $\phi(x,y)$. However, the authors in \cite{droulias2024near} simplify this integration by assuming that the $k$-content of the beam is concentrated within a narrow region around $k_r = k\sin{\theta_r}$. Furthermore, they exploit the phase curvature characteristics under the assumption that the focal distance $d_0$ exceeds the UPA size. Together, both considerations lead to simplified expressions for $k_z$ and $\phi(x,y)$
\begin{equation} \label{eq:k_z}
    k_z \approx \frac{3-\sec^2\theta_r}{2\cos{\theta_r}} + k_x \tan^3\theta_r - \frac{k_x^2}{2k\cos^3\theta_r} - \frac{k_y^2}{2k\cos{\theta_r}},
\end{equation}
\begin{equation}
\resizebox{1\columnwidth}{!}{%
    $\displaystyle \phi(x,y) \approx -k\sin{\theta_i}x +kd_0 \left(1-\frac{\tan^2\theta_r}{2}\right) + k\frac{(x\cos\theta_r - d_0 \tan\theta_r)^2}{2d_0} + k \frac{y^2}{2d_0}.$
}
\end{equation}
This enables the derivation of a closed-form solution for the radiated field $\mathbf{E}_r(x,y,z)$. Solving the integral in (\ref{eq:ASR}) yields an analytical expression for the squared magnitude of the propagated field at an arbitrary observation point $(x,y,z)$ as
\begin{equation} \label{eq:squared_field}
    |E_r(x,y,z)|^2 = |E_0|^2 \frac{w_{0,x}}{w_x(z)} \frac{w_{0,y}}{w_y(z)} e^{\left[-2\left(\frac{(x-z\tan\theta_r)^2}{w^2_x(z)}+\frac{y^2}{w^2_y(z)}\right)\right]},
\end{equation}
where $w_x(z)$ and $w_y(z)$ represent the beam radii along the $x$- and $y$-axis, respectively, at a given observation plane $z$
\begin{small}
\begin{equation} \label{eq:w_z}
w_{x}(z) = w_{0,x} \left[ \left( 1 - \frac{z}{d_{0} \cos \theta_{r}} \right)^{2} \right.
\left. + \left( \frac{z}{z_{R} \cos \theta_{r}} \right)^{2} \left( \frac{\cos \theta_{i}}{\cos \theta_{r}} \right)^{4} \right]^{1/2},
\end{equation}
\end{small}
\begin{equation}
w_{y}(z) = w_{0,y} \sqrt{\left( 1 - \frac{z}{d_{0} \cos \theta_{r}} \right)^{2} + \left( \frac{z}{z_{R} \cos \theta_{r}} \right)^{2}},
\end{equation}
with $w_{0,x} = w_{0} / \cos \theta_{i}$ and $w_{0,y} = w_{0}$ accounting for the elliptical footprint at the UPA. 
Evaluated along the steering direction ($x = r \sin \theta_{r}$, $y = 0$, $z = r \cos \theta_{r}$), the exponential argument in (\ref{eq:squared_field}) becomes zero. Consequently, the power density $S = \frac{|E|^2}{2\eta_0}$, with $\eta_0$ representing the free-space wave impedance, can be written in terms of $w_x(r)$ and $w_y(r)$ as
\begin{equation} \label{eq:power_density1}
S(r) = \frac{|E_{0}|^{2}}{2\eta_{0}} \frac{w_{0,x}}{w_{x}(r)} \frac{w_{0,y}}{w_{y}(r)}.
\end{equation}
Here, the transverse beam radii along the direction of propagation $r$ are given by
\begin{align}
w_{x}(r) &= w_{0,x} \Bigg[ \left( 1 - \frac{r}{d_{0}} \right)^{2}+ \left( \frac{r}{z_{R}} \right)^{2} \left( \frac{\cos \theta_{i}}{\cos \theta_{r}} \right)^{4} \Bigg]^{1/2}, \label{eq:w_x(r)} \\
w_{y}(r) &= w_{0,y} \sqrt{\left( 1 - \frac{r}{d_{0}} \right)^{2} + \left( \frac{r}{z_{R}} \right)^{2}}. \label{eq:w_y(r)}
\end{align}
Substituting (\ref{eq:w_x(r)}) and (\ref{eq:w_y(r)}) into (\ref{eq:power_density1}) yields $\frac{|E_0|^2}{2\eta_0\sqrt{\text{p}(r)}}$, where $p(r)$ is defined as
\begin{equation} \label{eq:p_r}
    \text{p}(r) = \left[\left(1-\frac{r}{d_0}\right)^2 + \left(\frac{r}{z_R}\gamma\right)^2\right] \times \left[\left(1-\frac{r}{d_0}\right)^2 + \left(\frac{r}{z_R}\right)^2\right],
\end{equation}
and the factor $\gamma = \frac{\cos^2{\theta_i}}{\cos^2{\theta_r}}$ relates the incidence angle $\theta_i$ to the steering $\theta_r$ angle. By expressing the propagation distance as $ r = \alpha d_0$, we obtain the fourth-degree polynomial
\begin{equation} \label{eq:polynomial_alpha}
    \text{p}(\alpha) = uv\alpha^4 - (2u+2v)\alpha^3 + (u+v+4)\alpha^2-4\alpha+1
\end{equation}
in which $u = 1 + \left(\frac{d_0}{z_R}\right)^2$ and $v = 1 + \left(\frac{d_0}{z_R} \gamma\right)^2$.

\begin{figure*}[!t]
    \centering
    \subfloat[]{
        \includegraphics[width=0.48\linewidth]{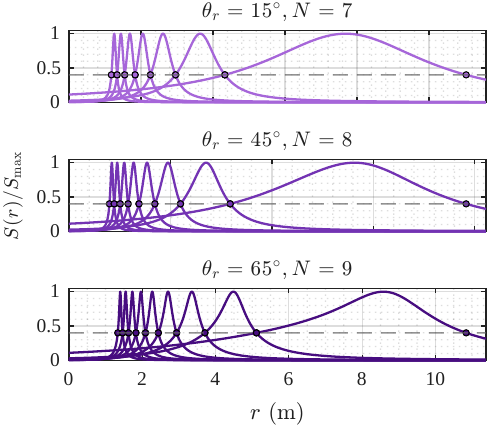}
        \label{fig:synthesis_non_uniform}
    }
    \hfil 
    \subfloat[]{
        \includegraphics[width=0.48\linewidth]{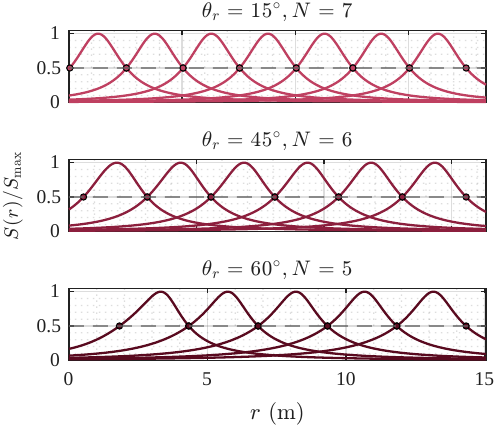}
        \label{fig:synthesis_uniform}
    }
    \caption{Resolving the radial domain under varying steering angles $\theta_r$ and inter-beam overlap $\rho_{\text{radial}}$ = 0.5.  (a) Non-uniform resolution model with a fixed beam radius $w_0$ = 0.2 m. (b) Uniform resolution model with a constant $w_{\text{FWHM}}$ = 2.5 m for all focal regions}
    \label{fig:synthesis_focal_regions}
\end{figure*}

To determine the position $r_{\text{peak}}$ corresponding to the maximum beam intensity $S_{\max}(r)$, we differentiate (\ref{eq:polynomial_alpha}) with respect to $\alpha$ and equate the result to zero. The analytical solution of $p'(\alpha) = 0$ is obtained by reducing the expression to a depressed cubic of the form $t^3 + mt + q = 0$, following Cardano's method. Applying the transformation $\alpha = t + \frac{u+v}{2uv}$ provides the coefficients $m$ and $q$ as
\begin{align*}
m &= \frac{2uv(u+v+1)-3(u^2+v^2)}{4u^2v^2}, \\
q &= \frac{(-u-v)^2(u+v-uv)}{4u^3v^3}. 
\end{align*}
The roots ($t_1, t_2, t_3)$ of the depressed cubic depend on its discriminant $\Delta_c$
\begin{equation} \label{delta_c}
\Delta_c = \frac{q^2}{4} + \frac{m^3}{27}.
\end{equation}
The condition $\Delta_c >0$ yields a single real root $t_1$, indicating that $S_{\max}(r)$ presents a unique maximum point at $r_{\text{peak}}$. According to Cardano's formula, this root is given by
\begin{equation}
    t_1 = \sqrt[3]{\frac{-q}{2}+\sqrt{\Delta_c}} + \sqrt[3]{\frac{-q}{2}-\sqrt{\Delta_c}}.
\end{equation}
Finally, to obtain $r_{\text{peak}}$, we substitute $t_1$ back into our initial variable definitions ($\alpha = t_1 + \frac{u+v}{2uv}$ and $\alpha= \frac{r_{\text{peak}}}{d_0}$), yielding the closed-form expression 
\begin{equation}\label{eq:r_peak}
\boxed{r_{\text{peak}} = d_0 \left( t_1 + \frac{u+v}{2uv} \right).}
\end{equation}

\section{Radial boundaries}
\label{sec:radial_boundaries}
The radial boundaries of the focal region are analytically derived following the approach in \cite{centritto2026analytical}, but considering the implications of a steered beam. To determinate the coordinates along the propagation direction $r$ where the peak power density $S_{\text{max}}$ decays to an inter-beam overlap parameter $\rho_{\text{radial}} \in [0,1)$, we solve the equation $S(r) = \rho_{\text{radial}} S_{\text{max}}$. Algebraic manipulation of this condition yields $\text{p(r)} = \frac{\text{p}(r_\text{max})}{\rho^2}$. Expressing $r$ via the scaling parameter $r = \alpha d_0$, we generate a fourth-degree polynomial $uv\alpha^4 - 2(u+v)\alpha^3+(u+v+4)\alpha^2-4\alpha+(1-\frac{\text{p}(r_{\text{peak}})}{\rho^2})$ that, using the prior $\alpha$ transformation, reduces to a depressed quartic $t^4 + mt^2 + qt + s = 0$. The resulting coefficients are

{\small
\begin{align*}
m &= \frac{2uv(u+v+4)-3(u^2+v^2)}{2u^2v^2},  \\
q &= \frac{uv(u+v)(u+v+4)-(u+v)^3-4u^2v^2}{u^3v^3},  \\
s &= \frac{-3(u+v)^4}{16u^4v^4} + \frac{(u+v)^2(u+v+4)}{4u^3v^3}-\frac{2(u+v)}{u^2v^2} + \frac{\rho-\text{p}(r_\text{peak})}{\rho^2uv}.
\end{align*}%
}
The depressed quartic is reorganized to express both sides as perfect squares following Ferrari's method. By introducing an auxiliary parameter $y$, the equation can be rewritten $((t^2+\frac{m}{2}) + y)^2 = 2yt^2 - qt + (y^2+my+\frac{m^2}{4}-s)$. A perfect square exists on the right-side only if its quadratic discriminant with respect to $t$ is zero. This requirement produces a cubic resolvent equation in $y$ of the form $y^3 + my^2 + \frac{ (m^2-4s)}{2}y - \frac{q^2}{8}$, which is solved using Cardano's formula to find the real root $y_1$ 
\begin{equation}
        y_1 = \sqrt[3]{\frac{-q_y}{2}+\sqrt{\Delta_{{c}_{y}}}} + \sqrt[3]{\frac{-q_y}{2}-\sqrt{\Delta_{{c}_{y}}}}.
\end{equation}
with $\Delta_{{c}_{y}} = \frac{q_y^2}{4} + \frac{m_y^2}{27}$, $m_y = -(m^2+s)$, and $q_y = \frac{ms}{3}-\frac{q^2}{8}-\frac{m^3}{108}$. Consequently, the depressed quartic can now be reduced into two separated quadratic equations
\begin{equation} \label{eq:two_quadratic_equation}
    t^2 + \frac{m}{2} + y_1 = \pm \left( \sqrt{2y_1}t - \frac{q}{2\sqrt{2y_1}}\right).
\end{equation}
By rearranging (\ref{eq:two_quadratic_equation}) and applying the quadratic formula, we obtain the four roots for $t$. The first two roots are given by $t_{1,2} = \sqrt{\frac{y_1}{2}} \pm \frac{1}{2}\sqrt{-2y_1 - 2m - q\sqrt{\frac{2}{y_1}}}$, and  
$t_{3,4} = -\sqrt{\frac{y_1}{2}} \pm \frac{1}{2}\sqrt{-2y_1 - 2m + q\sqrt{\frac{2}{y_1}}}$. Reversing the initial variable substitutions ($\alpha = t + \frac{(u+v)}{2uv}$ and $\alpha = \frac{r}{d_0}$) generates the four analytical solutions for $r$ are recovered as
\begin{equation}
r_i=d_0\left(t_i-\frac{(u+v)}{2uv}\right)\quad\text{for}\quad i\in\{1,2,3,4\}
\end{equation}
Through numerical evaluation, we find that the boundaries of the focal correspond to the real roots $r_1$ and $r_2$. This, the start $(r_\text{start})$ and end $(r_\text{end})$ positions of the focal region are analytically defined as

\begin{small}
\begin{equation}\label{eq:r_end}
\boxed{r_{end} = d_0 \Bigg[ \frac{u+v}{2uv} + \frac{1}{2}\sqrt{2y_1}\quad + \frac{1}{2}\sqrt{-2y_1 - 2m - q\sqrt{\frac{2}{y_1}}} \Bigg]} 
\end{equation}
\end{small}

\begin{small}
\begin{equation} \label{eq:r_start}
\boxed{r_{start} = d_0 \Bigg[ \frac{u+v}{2uv} + \frac{1}{2}\sqrt{2y_1}\quad - \frac{1}{2}\sqrt{-2y_1 - 2m - q\sqrt{\frac{2}{y_1}}} \Bigg]}
\end{equation}
\end{small}

Finally, the focal region extent $\Delta r_{\rho}$ is calculated as the difference between $r_{end}$ and $r_{start}$ yielding the closed-form solution

\begin{small}
\begin{equation} 
\boxed{
\begin{aligned}
\Delta r_{\rho} &= d_0 \Bigg[ -2\Bigg( \frac{2uv(u+v+4) - 3(u+v)^2}{2u^2v^2} + y_1 \\
&\quad + \frac{1}{\sqrt{2y_1}}\frac{-4u^2v^2 + uv(u+v)(u+v+4) - (u+v)^3}{u^3v^3} \Bigg) \Bigg]^{1/2}
\end{aligned}
}
\end{equation}
\end{small}

\subsection{Derivation of minimum focal distance $d_{0,\min}$ }

The $d_{0,\min}$ limit is analytically derived considering from the approximation of $k_z^{\text{approx}}$ made in (\ref{eq:k_z}). To determine the minimum focal distance at which this approximation accurately models the exact propagated field, we determine the maximum transverse spatial component wavenumber $k_x$ (setting $k_y=0$, as beam steering is restricted to the $xz$-plane) for which the approximation error $\Delta_{k_z} = \left\vert{} k_z^{\text{exact}} - k_z^{\text{approx}} \right\vert{}$ remains negligible. Specifically, $k_x$ is parametrized in terms of of the maximum tolerated $k$-component divergence $\theta_{\text{limit}}$ from the main propagation axis, yielding $k_x^{\max} = k \sin(\theta_{r} + \theta_{\text{limit}})$. Furthermore, the closed-form solution of $\vert{}\hat{E}^{\text{UPA}}(k_x, k_y)\vert{}$ is 
\begin{equation} \label{eq:mag_UPA}
\vert{} \hat{E}^{UPA} (k_x) \vert{} = \frac{E_0}{4\pi \sqrt{a_x \cdot a_y}} \exp\left( - \frac{(k_x - k \sin\theta_r)^2 }{\left(\frac{2 w_0 a_x }{\cos\theta_i} \right)^2}  \right),
\end{equation}
where $a_x = \sqrt{ \left( \frac{\cos^2\theta_i}{w_0^2} \right)^2 + \left( \frac{k \cos^2\theta_r}{2d_0} \right)^2 }$, and $a_y = \sqrt{ \left( \frac{1}{w_0^2} \right)^2 + \left( \frac{k}{2d_0} \right)^2 }$.
The spectral width of the steered beam, centered at $k_x = k \sin\theta_r$, is given by $\hat{w}_{0}=\frac{2w_{0}a_{x}}{\cos\theta_{i}}$, which places the beam's primary spectral support at $k \sin\theta_r + \hat{w}_0$. For the approximation to remain valid, the primary spectral $k$-content must be confined within $k_x^{\max}$, such that the inequality $k \sin\theta_r + \hat{w}_0 \leq k_x^{\max}$ must hold. Solving yields the closed-form expression for $d_{0,min}$

\begin{small}
\begin{equation} \label{eq:d0_min}
     \boxed{d_{0,\text{min}}  = \frac{2 z_R \cos^{2}\theta_{r}}{\cos\theta_{i}\sqrt{2kz_R \left(\sin(\theta_{r}+\theta_{\text{limit}}) - \sin(\theta_{r})\right)^{2} - 4\cos^{2}\theta_{i}}}}
\end{equation}
\end{small}

\subsection{Design of non-uniform and uniform focal regions in the radial domain}
To design $N$ contiguous focal regions with either non-uniform or uniform radial resolution centered at focal distances $d_{0,1},d_{0,1}...d_{0,N}$, the beams are distributed such that the upper radial boundary of the $(n-1)$-th focal region coincides with the lower boundary of the $n$-th region, satisfying the contiguity condition $r_{\text{end},n-1}$ = $r_{\text{start},n}$. Following the framework in \cite{centritto2026analytical}, our objective is to sequentially determine the adjacent focal distance $d_{0,n-1}$ from a known $d_{0,n}$. However, because the boundaries expressions in (\ref{eq:r_end}) and (\ref{eq:r_start}) depend on $d_{0,n-1}$ in a non-linear, nested manner through the roots of the underlying quartic polynomial, an explicit closed-form expression is analytically intractable. Consequently, we develop iterative numerical synthesis algorithms to compute the focal parameters of both resolutions profiles.

For the non-uniform resolution profile, the initial beam radius $w_0$ is held constant across all beams, reducing the synthesis to a one-dimensional root finding problem solved backward from the furthest focal region toward the radiating aperture. First, we obtain the focal distance of the outermost region $d_{0,N}$ by maximizing the lower boundary $r_{\text{start}}$ via bounded scalar optimization initialized from an initial random guess. Once $d_{0,N}$ and its corresponding starting boundary are established, the preceding focal distances $d_{0,n-1}$ are computed iteratively by finding the root of the residual function $f(d_{0,n-1}) = r_{\text{end}}$ starting with an initial random guess at each step. Additionally, to achieve uniform radial resolution across all regions, each beam must simultaneously satisfy the spatial contiguity condition and maintain a constant target focal extent $\Delta r_{\rho,\text{target}}$ mapped from a desired $w_{\text{FWHM}}$. Because this imposes two constraints per region, both the focal distance $d_{0,n-1}$ and the Rayleigh range $z_{R,n-1}$ must be jointly optimized. The synthesis is initialized at the furthest region by considering the maximum physical aperture size \cite{centritto2026analytical}, and numerically solving for $d_{0,N}$ such that $r_{\text{end}}(d_{0,N}, z_{R,\text{max}}) - r_{\text{start}}(d_{0,N}, z_{R,\text{max}}) = \Delta r_{\rho,\text{target}}$. Subsequently, for each adjacent inner region, the parameter vector $\mathbf{x}_{n-1} = [d_{0,n-1}, z_{R,n-1}]^T$ is obtained by solving the two-dimensional non-linear system $\mathbf{F}(\mathbf{x}_{n-1}) = [r_{\text{end}}(\mathbf{x}_{n-1}) - G_n, \, \Delta r(\mathbf{x}_{n-1}) - \Delta r_{\rho,\text{target}}]^T = \mathbf{0}$ via bounded non linear least-squares optimization. Both iterative procedures continue until the computed focal distance $d_{0,n-1}$ violates the limit $d_{0,n-1}<d_{0,\min}$, which remains constant for the non-uniform profile but varies dynamically with $z_{R,n-1}$ in the uniform resolution case.

To illustrate the behavior of the framework, Fig. \ref{fig:synthesis_focal_regions} shows the normalized radial power density $S(r)/S_{\max}$ profiles evaluated for steering scenarios at $f = 150\text{ GHz}$, $\theta_i = 0^\circ$, $\rho=0.5$ and $\theta_{\text{limit}} = 4^\circ$. For the non-uniform resolution scenario in Fig. \ref{fig:synthesis_non_uniform}, holding $w_0 = 0.2$ m constant causes the radial extent $\Delta r_\rho$ to progressively narrow towards the aperture. As $\theta_r$ increases $\{15^\circ, 45^\circ, 65^\circ\}$, the minimum focal distance limit lowers to yield $N = 7, 8,$ and $9$ contiguous regions, respectively. Conversely, the uniform resolution design in Fig. \ref{fig:synthesis_uniform} maintains a constant focal extent of $w_{\text{FWHM}} = 2.5\text{ m}$ by jointly optimizing the focal distance and the beam radius. In this regime, increasing the steering angle across $\theta_r \in \{15^\circ, 45^\circ, 60^\circ\}$ reduces the synthesizable focal regions from $N = 7$ to $6$ and $5$, as preserving a constant $w_{\text{FWHM}}$ under larger $\theta_r$ requires higher $w_{0,n}$ values, which elevates the dynamic $d_{0,\min}$ and restricts focusing near the array.

\section{Angular boundaries}
\label{sec:angular-boundaries}
To determine the angular boundaries $\theta_{\text{start}}$ and $\theta_{\text{end}}$ at which the maximum power density $S_{\max}$ decays a fraction $\rho_{\text{angular}}$, the power density in $xz$-plane is evaluated as a function of the observation angle $\theta$ along the constant-radius arc $r = r_{peak}$ arc. Consequently, we parametrize the spatial coordinates as $x = r_{\text{peak}} \sin\theta$ and $z = r_{\text{peak}} \cos\theta$, and substituting them into the transversal exponent in (\ref{eq:squared_field}) yields $x - z \tan\theta_r = \frac{r \sin(\theta - \theta_r)}{\cos\theta_r}$.  Defining the angular deviation from the steering direction $\theta_r$ as $\delta_\theta = \theta - \theta_r$ and enforcing the threshold condition $S(\theta) = \rho_{\text{angular}} S_{\max}$, we obtain
\begin{align*}
    \exp{\left[-2\left(\frac{r_{\text{peak}} \sin\delta_\theta}{w_{x,\text{peak}} \cos\theta_r}\right)^2\right]} = \rho
\end{align*}
which can be inverted to express $\delta_\theta$ in closed form as
\begin{equation}\label{eq:delta_theta}
    \boxed{\delta_\theta = \arcsin\left( \frac{w_{x,\text{peak}} \cos\theta_r}{r_{\text{peak}}} \sqrt{\frac{-\ln(\rho)}{2}} \right)}
\end{equation}
From this deviation, the lower and upper angular boundaries of the focal region and the resulting angular extent $\Delta\theta_\rho$ are given, respectively, by
\begin{equation}
\boxed{\theta_{\text{start}} = \theta_r - \delta_{\theta}, \quad \theta_{\text{end}} = \theta_r + \delta_{\theta}, \quad \Delta\theta_\rho = \theta_{\text{end}} - \theta_{\text{start}} = 2\delta_\theta.}
\end{equation}
\begin{figure}
        \centering
        \includegraphics[width=1\linewidth]{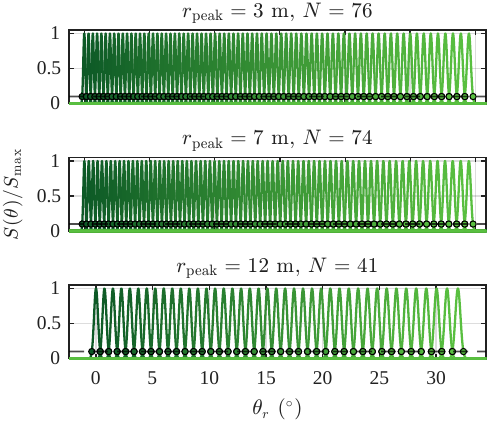}
        \caption{Resolving the angular domain at multiple radial distances $r_{\text{peak}}$, considering beam radius of $w_0$ = 0.15 m and inter-beam overlap $\rho_{\text{angular}}$ = 0.1.}
        \label{fig:synthes_angular_domain}
    \end{figure}
These boundaries establish a direct framework for contiguous angular beam design: for a fixed radial distance $r_{\text{peak}}$ shared across all beams, adjacent focal regions can be concatenated along the angular domain $\theta$ such that their profiles intersect at the inter-beam overlap $\rho_{\text{angular}}$, satisfying the angular contiguity condition $\theta_{\text{end,n-1}} = \theta_{\text{start,n}}$.  

\begin{figure*}[!t]
    \centering
    \subfloat[]{
        \includegraphics[width=0.3\linewidth]{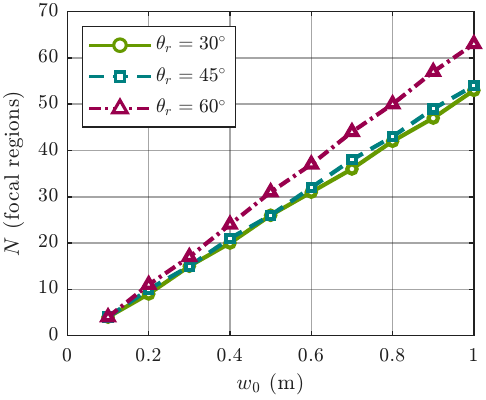}
        \label{fig:non_uniform_extent_resolution_capabilities}
    }
    \hfil 
    \subfloat[]{
        \includegraphics[width=0.3\linewidth]{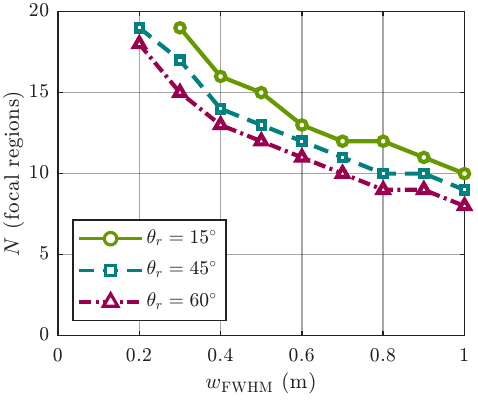}
        \label{fig:uniform_extent_resolution_capabilities}
    }
    \hfil
    \subfloat[]{
        \includegraphics[width=0.3\linewidth]{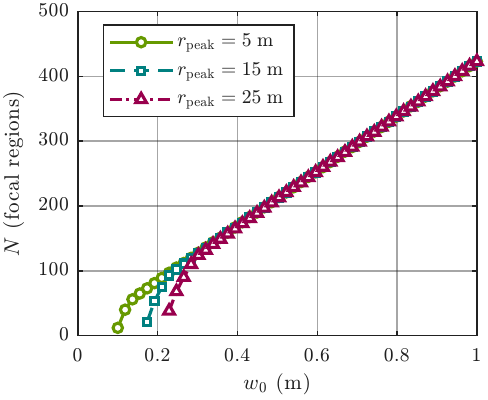}
        \label{fig:angular_framework_resolution_capabilities}
    }
    \caption{Resolution capabilities in the radial and angular domain with $f = 150$ GHz (a) Non-uniform resolution model, illustrating the number of focal regions $N$ with respect to the beam radius $w_0$ for different $\theta_r$. (b) Uniform resolution model, where $N$ is detailed as a function of the $w_{\mathrm{FWHM}}$ for different steering angles $\theta_r$. (c) Angle resolution configuration, showing $N$ as a function of the beam radius $w_0$ for various $r_{\text{peak}}$.}
    \label{fig:resolution_capabilities}
\end{figure*}

\subsection{Derivation of maximum steering angle $\theta_{\max}$}

In angular design framework, maintaining a constant target peak distance $r_{\text{peak,target}}$ across all focal regions imposes a maximum on the steering angle. As $\theta_r$ increases, the maximum achievable peak distance $r_{\text{peak}}^*$ progressively decreases. Therefore, realizing a valid focal region configuration at a given steering angle requires satisfying the condition $r_{\text{peak,target}} \leq r_{\text{peak}}^*(\theta_r)$. To determine $r_{\text{peak}}^*$ analytically from the power density polynomial $\text{p}(r)$ in (\ref{eq:p_r}), two conditions must be satisfied simultaneously. First, by definition, the radial peak occurs where the first partial derivative vanishes, $\frac{\partial p}{\partial r} = 0$. Second, treating $r_{\text{peak}}$ as an implicit function of $d_0$, the global maximum $r_{\text{peak}}^*$ occurs at the stationary point $\frac{d r_{\text{peak}}}{d d_0} = 0$. Differentiating $\frac{\partial p}{\partial r} = 0$ implicitly with respect to $d_0$ yields $$\frac{\partial^2 p}{\partial r^2} \frac{d r_{\text{peak}}}{d d_0} + \frac{\partial^2 p}{\partial r \partial d_0} = 0.$$
Enforcing $\frac{d r_{\text{peak}}}{d d_0} = 0$ causes the first term to vanish, reducing the system to the mixed partial derivative condition $\frac{\partial^2 p}{\partial r \partial d_0} = 0$. Introducing the dimensionless ratio $x = d_0 / r_{\text{peak}}^*$ (such that $d_0 = x r_{\text{peak}}^*$) allows $\frac{\partial^2 p}{\partial r \partial d_0} = 0$ to be simplified into the algebraic relation
$$(r_{\text{peak}}^*)^2 x^2 (3x - 4)(\gamma^2 + 1) + 2z_R^2 (x - 1)^2 (x - 4) = 0.$$
Solving this relation for $r_{\text{peak}}^*$ yields the closed-form expression
\begin{equation} \label{eq:r_max_peak}
    \boxed{r_{\text{peak}}^* = z_R \frac{x - 1}{x} \sqrt{\frac{2(4 - x)}{(3x - 4)(\gamma^2 + 1)}}}.
\end{equation}
To obtain the unknown ratio $x$, both $r = r_{\text{peak}}^*$ and $d_0 = x r_{\text{peak}}^*$ are substituted back into the primary condition $\frac{\partial p}{\partial r} = 0$. This substitution eliminates the radial variable and reduces the condition to a cubic polynomial $P x^3 + Q x^2 + R x + S = 0,$ where $P = 3\gamma^4 + 2\gamma^2 + 3$, $Q = -13\gamma^4 + 10\gamma^2 - 13$, $R = 24(\gamma^2 - 1)^2$, and $S = -16(\gamma^2 - 1)^2$. Solving this cubic equation for its valid root and replacing it into (\ref{eq:r_max_peak}) determines the maximum achievable peak distance. Consequently, the maximum allowable steering angle $\theta_{\max}$ is defined as the angle beyond which the condition $r_{\text{peak,target}} \leq r_{\text{peak}}^*(\theta_r)$ is violated.

\subsection{Design of contiguous focal regions in the angular domain}
To generate $N$ contiguous focal regions along a constant-radius arc with a fixed beam radius $w_0$, both the steering angle $\theta_{r,n}$ and the focal distance $d_{0,n}$ are sequentially determined for each beam. The iterative procedure begins by initializing an anchor beam ($n = 1$) at a prescribed steering angle $\theta_{r,1}$. After verifying that the the target $r_{\text{peak}}$ satisfies $r_{\text{peak}} \leq r_{\text{peak}}^*(\theta_{r,1})$, $d_{0,1}$ is obtained via one-dimensional root-finding such that $r_{\text{peak}}(d_{0,1}, \theta_{r,1}) - r_{\\text{peak}} = 0$, which establishes the angular boundaries $\theta_{\text{start},1}$ and $\theta_{\text{end},1}$. For each subsequent beam, maintaining a constant radial peak while enforcing angular contiguity $\theta_{\text{start},n} = \theta_{\text{end},n-1}$ requires jointly solving the two-dimensional non linear system $\mathbf{F}(\theta_{r,n}, d_{0,n}) = [r_{\text{peak}}(d_{0,n}, \theta_{r,n}) - r_{\text{peak}}, \, \theta_{\text{start}}(d_{0,n}, \theta_{r,n}) - \theta_{\text{end},n-1}]^T = \mathbf{0}$ from an initial guess. This angular recursion continues until the steering angle exceeds the maximum system limit ($\theta_{r,n} > \theta_{\text{peak}}$) or the desired peak radius violates the analytical condition $r_{\text{peak}} > r_{\text{peak}}^*(\theta_{r,n})$.

This angular synthesis is shown in Fig. \ref{fig:synthes_angular_domain}, where the normalized angular power density profiles $S(\theta)/S_{\max}$ evaluated at $f = 120\text{ GHz}$ for a beam radius of $w_0 = 0.15\text{ m}$, normal incidence ($\theta_i = 0^\circ$), and inter-beam overlap of $\rho_{\text{angular}} = 0.1$. Comparing the three radial distances $r_{\text{peak}} \in \{3, 7, 12\}\text{ m}$ reveals two important behaviors. First, within each profile, as $\theta_{r,n}$ increases towards more oblique angle, $\gamma_n$ expands the transverse beam radius $w_{x,\text{peak}}$, causing the angular extent $\Delta\theta_{\rho,n}$ to progressively widen. Second, increasing from $r_{\max} = 3\text{ m}$ to $7\text{ m}$ and $12\text{ m}$ reduces the number of focal regions from $N = 76$ to $N = 74$ and $N = 41$, respectively. This reduction happens because a larger $r_{\text{peak}}$ reaches the analytical ceiling $r_{\text{peak}}^*(\theta_r)$ at a smaller angle $\theta_{\max}$, terminating the angular chain of focal regions earlier.

\section{Resolution capabilities for radial and angular domain}
\label{sec:resolution_analysis}

\begin{figure*}[!t]
    \centering
    \subfloat[]{\includegraphics[width=1\columnwidth]{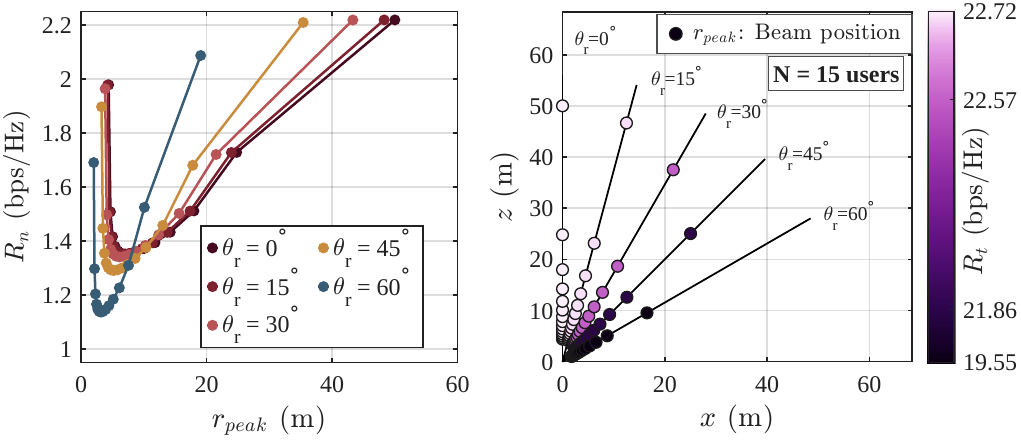}\label{fig:non_uniform_performance}}
    \hfil
    \subfloat[]{\includegraphics[width=1\columnwidth]{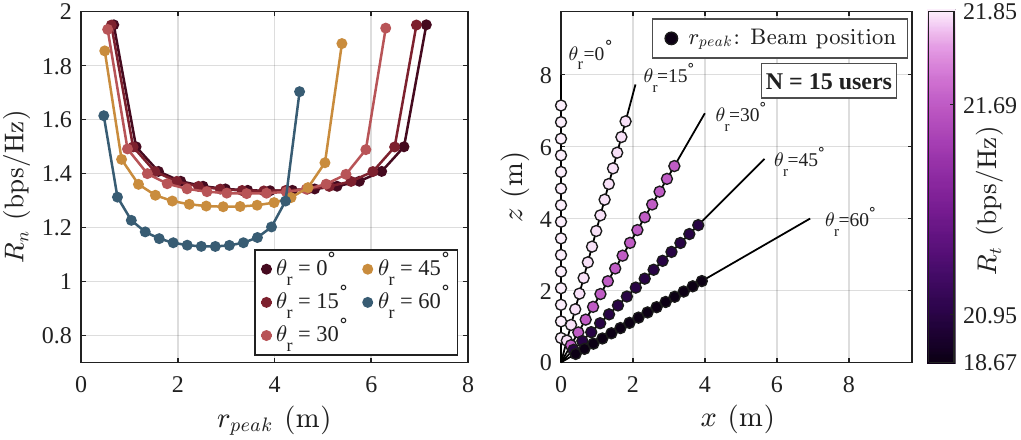}\label{fig:uniform_performance}}
    \caption{Per-user spectral efficiency $R_n$ and radial beam locations ($r_{\text{peak}}$) in the $xz$-plane alongside the system sum-rate $R_t$ across various steering angle for a fixed user capacity of $N = 15$ inter-beam radial overlap $\rho = 0.5$. (a) Non-uniform radial resolution. (b) Uniform radial resolution.}
    \label{fig:non-uniform_uniform_performance_assessment}
\end{figure*}

\begin{figure*}[!t]
    \centering
    \subfloat[]{\includegraphics[width=1\columnwidth]{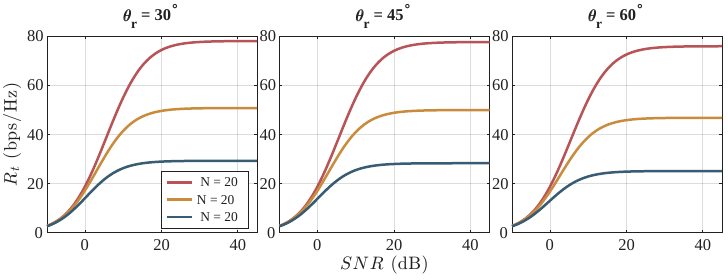}\label{fig:N_fixed_snr_vs_sum_rate}}
    \hfil
    \subfloat[]{\includegraphics[width=1\columnwidth]{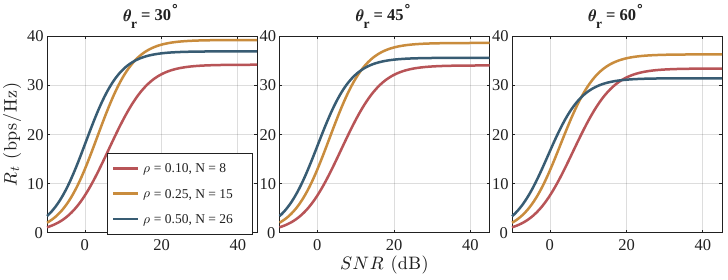}\label{fig:N_var_snr_vs_sum_rate}} 
    \caption{System sum-rate $R_t$ as function of SNR across steering angles $\theta_r \in \{30^\circ, 45^\circ, 60^\circ\}$ and radial inter-beam overlap parameters $\rho \in \{0.10, 0.25, 0.50\}$ for the non-uniform radial resolution framework. (a) Fixed user capacity ($N = 20$). (b) Variable user capacity ($N \in \{8, 15, 26\}$.}
    \label{fig:sumrate_vs_SNR}
\end{figure*}

The user-capacity of both radial and angular resolution framework for locating contiguous focal region is studied in detail. In the non-uniform resolution model, the fixed beam radius $w_0$ is the primary parameter dictating the system capability. As $w_0$ expands, the Rayleigh range $z_R$ increases quadratically, thereby extending the near-field boundary; consequently, the outermost focal distance $d_{0,N}$ that maximizes $r_{start}$, is shifted further outward. This yields an synthesizable range, resulting in a proportionally higher number of contiguous focal regions $N$, as illustrated in Fig. \ref{fig:non_uniform_extent_resolution_capabilities}. For instance, at a fixed steering angle of $\theta_r = 30^\circ$, expanding the beam radius from $w_0 = 0.25$ m to $w_0 = 1$ m scales the capacity from approximately 12 to 60 focal regions. Furthermore, while increasing the steering angle $\theta_r$ pulls this outermost beam $d_{0,N}$ closer to the array, the total number of focal regions $N$ still grows. This is explained by the relationship between $d_{0,\min}$ and $\theta_r$ in (\ref{eq:d0_min}):as $\theta_r$ expands, $d_{0,\min}$ shifts significantly closer to the aperture. As demonstrated in Fig. \ref{fig:non_uniform_extent_resolution_capabilities}, this inward shift of $d_{0,\min}$ dictates the overall scaling of the accessible radial domain. The expansion of the coverage region near the array outpaces the range reduction of the outermost beam, increasing the number of supportable users. This effect becomes more pronounced at larger apertures; at $w_0 = 0.8$ m, augmenting $\theta_r = 30^\circ$ to $60^\circ$ increase the supported regions from 40 to 50.

Conversely, the system is characterized by a constant focal resolution of $w_{\text{FWHM}}$ across all regions for the uniform framework, tailoring distinct $w_0$ per beam. As shown in Fig. \ref{fig:uniform_extent_resolution_capabilities}, expanding $w_{\text{FWHM}}$ reduces the total user capacity. For example, at a $30^\circ$ steering angle, increasing $w_{\text{FWHM}}$ from 0.3 m to 1.0 m reduces the capacity from a peak of 18 down to 10 regions. Additionally, the total capacity also decreases as $\theta_r$ increases. At a fixed focal extent of 0.4 m, $N$ drops from 16 regions at $\theta_r = 30^\circ$ to 13 regions at $60^\circ$. This reduction arises from the constraints imposed in the outermost and inner boundaries of the available range domain. At the upper boundary, larger $\theta_r$ pull the furthest focal region (constrained by $w_{0,\max}$) closer to the array; while, at the lower boundary, the focal spot is positioned far enough from the array to physically accommodate the required $w_{\text{FWHM}}$; 
The minimum focal distance $d_{0,\min}$ does define an important restriction on this resolution framework: the minimum input $w_{\text{FWHM,min}}$ that can be synthesized without breaking the paraxial constraint (\ref{eq:k_z}). This translates into the finest resolution that can be achieved, and it is obtained by evaluating $\Delta_r$ at $d_{0,\min}$ with $\rho = 0.5$. In Fig. \ref{fig:uniform_extent_resolution_capabilities}, we observe that smaller $\theta_r$ impose a larger minimum allowable focal extent $w_{\text{FWHM,min}}$. 

For the angular domain framework, the beam radius $w0$ and the constant arc defined by $r_{\text{peak}}$ are the main parameters that fix the resolution of the system. An increase in the Rayleigh Range $z_R$, associated to higher values of $w_0$, extends the maximum range $r_{\max}^*$ further from the array for all steering angles. As illustrated in Fig. \ref{fig:angular_framework_resolution_capabilities}, this ultimately yields a higher user-capacity as $w_0$ increases.Furthermore, different radial distances $r_{peak}$  yields the same number of users for most $w_0$ values. To understand the independence between the $N$ and $r_{peak}$, we must consider (\ref{eq:delta_theta}), where the angular deviation $\delta_\theta$ from the steering angle is determined by the ratio $\frac{w_{x,peak}}{r_{peak}}$. At the focal peak $r_{peak} \approx d_0$, the term \mbox{$(1 - \frac{r_{peak}}{d_0})^2$} in (\ref{eq:w_x(r)}) approaches zero, therefore \mbox{$w_{x,peak} \approx w_{0,x} \left(\frac{r_{peak}}{z_R \cos \theta_r}\right) \left(\frac{\cos \theta_i}{\cos \theta_r}\right)^2$}. If this $w_{x,peak}$ approximation is substituted back into (\ref{eq:delta_theta}), the radial depth $r_{peak}$ cancel out. Consequently, the angular resolution (that directly influences $N$) is independent to different values of $r_{peak}$. Due to the fine angular resolution inherent to the angular framework, narrow angular extent allows the synthesis of over 400 contiguous focal regions along a constant arc. This constitutes a significative capacity increase compared to the radial domain, which supports a maximum of 60 focal regions for a $1$ m beam radius in the non-uniform scenario.

\section{Per-user and sum-rate spectral efficiency assessment}
\label{sec:system_performance}
\begin{figure*}[!t]
    \centering
    \subfloat[]{
        \includegraphics[width=0.55\linewidth]{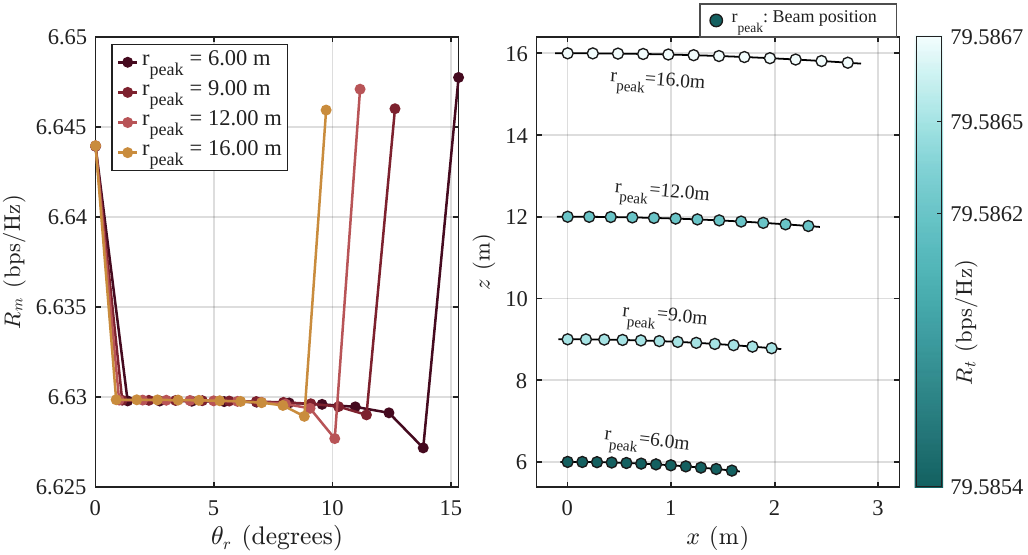}         \label{fig:per-user-efficiency-angular}}
    \subfloat[]{
        \includegraphics[width=0.37\linewidth]{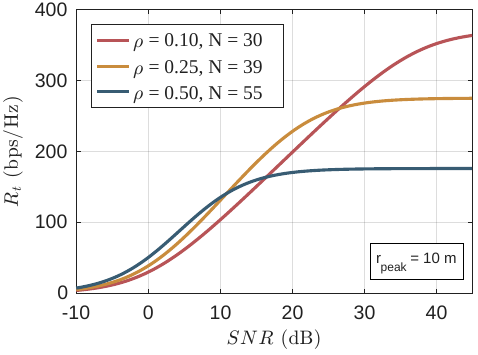} \label{fig:sum-rate-efficiency-angular}}
        
    \caption{System performance assessment for angular framework. (a) per-user and total sum-rate performance at different $r_{peak}$ values at fixed \textbf{N = 12 users}. (c) Sum-rate variation vs. SNR for variable user scenario.}
    \label{fig:angular_performance_analysis}
\end{figure*}
The multiple-access performance of the $N$ focal regions synthesized along the $xz$- plane is evaluated in terms of spectral efficiency. This metric is governed by the signal-to-interference-plus-noise ratio (SINR) measured at the peak intensity $r_{\text{peak}}$ of each focal region. For the $n$-th beam, the SINR is given by $\text{SINR}_n  = \ \frac{\text{SNR}}{\alpha_n \cdot \text{SNR} + 1}$, where $\text{SNR} = P_n / P_\nu$. Here, $P_n$ represents the intended peak received power, and $P_\nu$ is the additive noise power. The parameter $\alpha_n = \sum_{l \neq n} P_{l,n} / P_n$ characterizes inter-beam interference, with $P_{l,n}$ denoting the leakage power from beam $l$-th beam observed at the focal point of the $n$-th beam.  Using this SINR formulation, the achievable rate for the $n$-th beam is computed as $R_n = \log_2(1 + \text{SINR}_n)$; consequently, the system sum-rate $R_{\text{t}}$ is obtained by aggregating the individual rates across all $N$ beams, such that $R_{\text{t}} = \sum_n^N R_n$

Fig.~\ref{fig:non-uniform_uniform_performance_assessment} compares the performance of uniform versus non-uniform radial resolution configurations across steering angles $\theta_r$ for a fixed capacity of $N = 15$ users with an inter-beam overlap of $\rho = 0.5$. In both configurations, the outermost and innermost radial boundaries achieve the highest individual rates because they experience interference from only one adjacent side. With non-uniform radial resolution, dense beams clustering near the array exacerbates mutual interference, producing a sharp and localized rate drop immediately beyond the innermost beam, as shown in Fig. \ref{fig:non_uniform_performance}. In contrast, Fig. \ref{fig:uniform_performance} depicts that uniform spacing maintains constant inter-beam separation, yielding a broader and more balanced per-beam rate distribution across central users. For both scenarios, increasing the steering angle $\theta_r$ degrades the performance of the focal regions, with the most severe reduction occurring at $\theta_r = 60^\circ$. This degradation is attributed to the radial compression of the focal regions required to accommodate a constant number of users. Increasing the steering angle restricts the maximum radial extent, pulling the furthest beam significantly closer to the array. The resulting reduction in synthetizable radial space forces a denser user distribution, thereby elevating inter-beam interference and degrading the total sum-rate of the system.

To analyze the influence of the inter-beam overlap parameter $\rho_{\text{radial}}$ on the performance, Fig. \ref{fig:sumrate_vs_SNR} illustrates how the total sum-rate changes as a function of the SNR at different $\rho$ values and $\theta_r$ angles for the non-uniform resolution framework. As illustrated in Fig. \ref{fig:N_fixed_snr_vs_sum_rate}, when comparing systems with an identical number of users, the sum-rate is maximized by the configuration that experiences the lowest inter-beam interference. Consequently, across all steering scenarios, the system with minimal overlap ($\rho = 0.1$) yields the highest capacity, approaching 80 bps/Hz for $\theta_r = 30^{\circ}$ and $45^{\circ}$; while the configuration ($\rho = 0.5$) with greater overlap ($\rho = 0.5$) demonstrates the poorest performance, saturating below 30 bps/Hz for $\theta_r = 30^{\circ}$. Additionally, the shape of the resulting curves in Fig. \ref{fig:N_fixed_snr_vs_sum_rate} demonstrates the system's transition from a noise-limited to an interference-limited regime as signal power increases. At lower SNR values (below 0 dB), the curves for all overlap scenarios are closely packed together, showing that performance is primarily restricted by noise rather than the spatial interference dictated by $\rho$. As the SNR increases from 0 dB to approximately 20 dB, the sum-rate expands steeply and the curves diverge significantly based on their respective inter-beam overlap values. Beyond approximately 20 to 30 dB, the sum-rate for all configurations flattens into a plateau. By allowing a different number of users between $\rho$ configurations in Fig. \ref{fig:N_var_snr_vs_sum_rate}, a distinct performance dynamic emerges. The noise-limited regime depicts that the sum-rate is amplified by increasing $N$; consequently, across $\theta_r$ range, the highly overlapped configuration ($\rho = 0.5$, $N = 26$) maximizes the sum-rate at lower SNR values. However, as the SNR increases and the system transitions into an interference-limited state, the severe inter-beam interference caused by densely packing 26 users stops further capacity scaling. This causes the performance curves to intersect; the intermediate configuration of $\rho = 0.25$ ($N = 15$) achieves a superior balance between user capacity and interference mitigation, ultimately yielding the highest sum-rate at the high-SNR plateau across all evaluated $\theta_r$. At the most extreme steering angle ($\theta_r = 60^\circ$), the required radial compression degrades the dense $\rho=0.5$ system, causing its saturation capacity to fall below the configuration with less supported users ($\rho=0.10$, $N=8$). These findings indicate that maximizing the sum-rate in high-SNR regime requires an optimal balance between the increasing the number of focal areas and mitigating inter-beam interference. A detailed analysis of this trade-off, evaluated as a function of the tunable inter-beam overlap parameter $\rho$, is provided in Appendix \ref{sec:appendix_A}.

The performance of the angular system framework is evaluated in Fig. \ref{fig:angular_performance_analysis}, considering different radial peak positions ($r_{\text{peak}}$) with an inter-beam parameter $\rho_{\text{angular}} = 0.1$. Unlike radial configurations, where focal regions share a common propagation axis, the angular separation of beams in this scenario significantly mitigates the inter-beam interference. As a result, this approach yields a higher per-user spectral efficiencies, and consequently, an elevated sum-rate compared to radial systems. As illustrated in Fig. \ref{fig:sum-rate-efficiency-angular}, the fundamental trade-off with $\rho$ also holds in the angular domain: dense user packing ($\rho = 0.5$) maximizes throughput at low SNR, while moderate overlap ($\rho = 0.1$) dominates the interference-limited high-SNR regime by maintaining adequate beam separation.

\begin{figure}[!t]
    \centering
    \subfloat[]{\includegraphics[width=0.48\columnwidth]{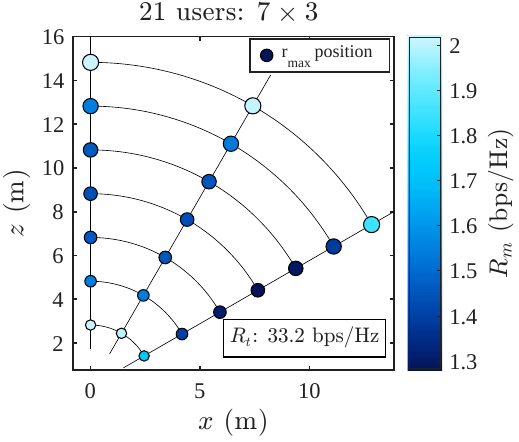}%
    \label{fig:mixed_system_panel_a}}
    \hfil
    \subfloat[]{\includegraphics[width=0.48\columnwidth]{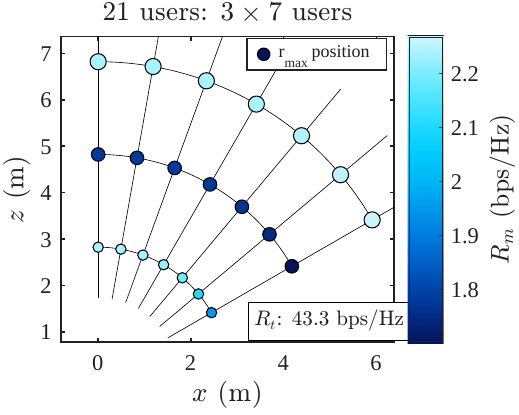}%
    \label{fig:mixed_system_panel_b}}
    
    \vspace{1em} 
    
    \subfloat[]{\includegraphics[width=0.48\columnwidth]{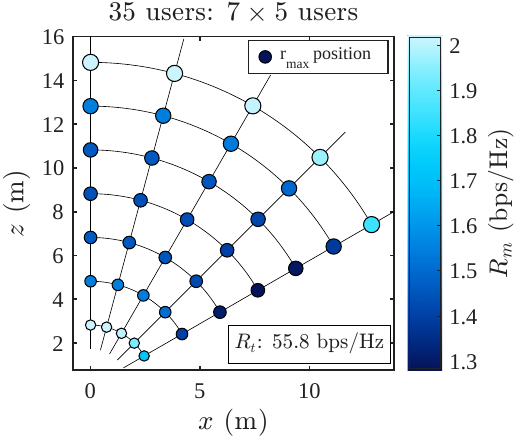}%
    \label{fig:mixed_system_panel_c}}
    \hfil
    \subfloat[]{\includegraphics[width=0.48\columnwidth]{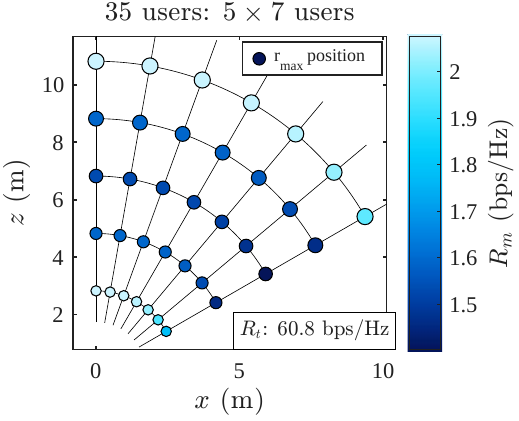}%
    \label{fig:mixed_system_panel_d}}
    
    \caption{Mixed angular-radial topologies per-user spectral efficiency comparison for different geometries and number of users $N$.}
    \label{fig:mixed_system}
\end{figure}

Building upon the interference reduction and superior spectral efficiency observed in the isolated angular analysis, a mixed system can be established to distribute focal regions jointly across both the radial and angular domains within the $xz-$ steering plane. To configure this spatial grid, $N_{radial}$ focal regions are first synthesized along the normal radial axis with $\theta_r = 0^\circ$ utilizing the uniform resolution framework. The equally spaced peaks of these regions establish a series of constant $r_{peak}$ arcs. Along these respective arcs, $N_{angular}$ users are then positioned across the angular domain, spanning from $0^\circ$ up to $60^\circ$, while enforcing a constant-width constraint to maintain uniform radial resolution. Fig. \ref{fig:mixed_system} evaluates the per-user $R_n$ of this topology by comparing different spatial geometries under fixed user capacities. For a 21-user network, arranging the focal regions into a configuration of 7 radial users by 3 angular users yields a sum-rate of 33.2 bps/Hz. By inverting this geometry to accommodate 3 radial users and 7 angular users, the system's sum-rate increases to 43.2 bps/Hz. Corroborating earlier findings from isolated domains, per-user efficiency improves when focal regions are distributed broadly across the angular space rather than aligned along the radial axis. This efficiency trend persists as the system grows. In a 35-user scenario, a geometry prioritizing radial resolution (7 radial by 5 angular users) achieves a sum rate of 55.8 bps/Hz, whereas prioritizing angular spread (5 radial by 7 angular users) increases the sum rate to 60.7 bps/Hz. The result establishes that for an equivalent user capacity, allocating a higher number of beams to the angular domain consistently maximizes overall spectral efficiency. 

\section{Conclusions}
\label{sec:conclu}
This paper introduces a resolution framework based on the ASR approach to characterize contiguous focal regions across both the radial and angular domains for $xz$-steered near-field multi-user systems. By extending the inter-beam overlap parameter $\rho$ as a tunable design variable in both domains, we derived closed-form expressions for the radial boundaries ($r_{\text{start}}$, $r_{\text{end}}$) and focal extent ($\Delta r_\rho$), alongside the angular boundaries ($\theta_{\text{start}}$, $\theta_{\text{end}}$) and extent ($\Delta\theta_\rho$) along a constant-radius arc $r_{\text{peak}}$. Furthermore, we analytically established the steering-dependent spatial limits, specifically the minimum focal distance $d_{0,\text{min}}$ and the maximum steering angle $\theta_{\text{max}}$, and developed numerical synthesis methodologies to construct contiguous focal regions either with uniform or non-uniform resolution for the radial domain. System-level evaluation demonstrated that while uniform radial spacing yields a more balanced per-user rate distribution than non-uniform, the angular domain supports a significantly higher number of contiguous focal regions, because angular resolution is independent of $r_{\text{peak}}$ and experiences lower inter-beam interference. Moreover, assessments of mixed angular-radial topologies confirmed that, for an equivalent user capacity, allocating a higher number of users to the angular domain rather than the radial maximizes overall spectral efficiency, underscoring the importance of a tunable design tool $\rho$ in the range domain to balance user-capacity against inter-user interference. Overall, these findings offer practical design insights for next-generation network architectures, providing a tractable methodology to control spatial resolution and optimize spectral efficiency in multi-user deployments.


\appendices

\section{$\rho_{\text{opt}}$ for maximum capacity along the steering plane.}
\label{sec:appendix_A}
\begin{figure}
    \centering
    \subfloat[]{\includegraphics[width=0.45\columnwidth]{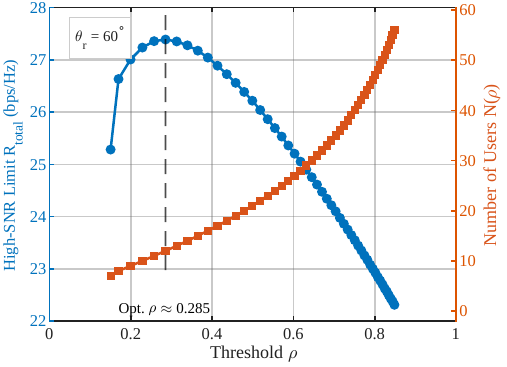}\label{fig:optimal_rho_evaluation}}
    \hfil
    \subfloat[]{\includegraphics[width=0.45\columnwidth]{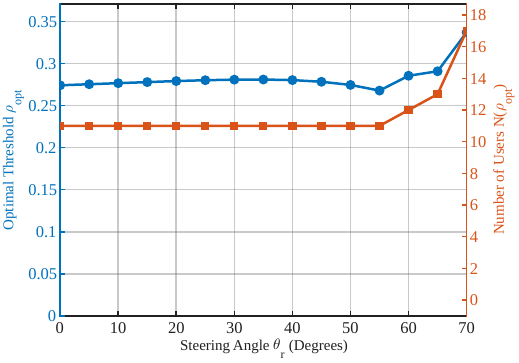}\label{fig:rho_steering}}
    \caption{Optimum $\rho_{\text{radial}}$ value analysis for steering in the $xz$ plane. (a) Sum-rate at high-SNR regime for varying values of overlap and identification of optimal $\rho$ configuration, at \mbox{$\theta_r = 60^\circ$}. (b) Variation of $\rho_{\text{opt}}$ for different steering angles $\theta_r$ }
\end{figure}

Figure \ref{fig:N_var_snr_vs_sum_rate} illustrates a fundamental tradeoff that emerges in the high-SNR regime of sum-rate $R_t$: while increasing the number of users $N$ initially enhances the sum-rate, dense radial inter-beam overlap eventually induces severe inter-beam interference that degrades overall performance, thereby motivating the identification of an optimal $N$ where the $R_t$ is maximized. The balance is governed by the overlap parameter $\rho$, as it simultaneously dictates the system's user capacity and the resulting interference levels. To explicitly capture this behavior, Fig. \ref{fig:optimal_rho_evaluation} evaluates the high-SNR sum-rate limit alongside the corresponding user capacity $N(\rho)$ as a function of $\rho_{\text{radial}}$ at $\theta_r = 60^\circ$. As $\rho$ increases, the system  accommodates a larger number of users. The sum-rate initially rises due to the capacity gains provided by these additional users, but eventually reaches a maximum and declines as inter-beam interference overtakes these gains. At this specific angle, system performance is maximized at an optimal threshold of $\rho_{opt} \approx 0.285$. Specifically, Fig. \ref{fig:rho_steering} demonstrates the variation of this optimal threshold across different steering angles. For angles up to $55^\circ$, the required $\rho_{opt}$ remains highly stable at approximately 0.285, yielding an optimal configuration of 11 users. However, at extreme oblique angles ($\theta_r > 55^\circ$), the radial compression of the focal regions forces the system to increase both the optimal inter-beam overlap threshold and the corresponding user capacity to maintain maximum spectral efficiency. Consequently, these findings validate the utility of the proposed analytical framework. By treating the spatial overlap $\rho$ as a tunable design variable the system can deterministically balance multi-user gains against interference penalties, ensuring optimized capacity across a wide range of steering scenarios.

\section*{Acknowledgments}
This work was supported by the European Union (HORIZON-MSCA-2023-DN-01), Grant Agreement 101169044 - TeraWireless.
\bibliographystyle{IEEEtran}
\bibliography{bib}

\begin{IEEEbiography}[{\includegraphics[width=1in,height=1.25in,clip,keepaspectratio]{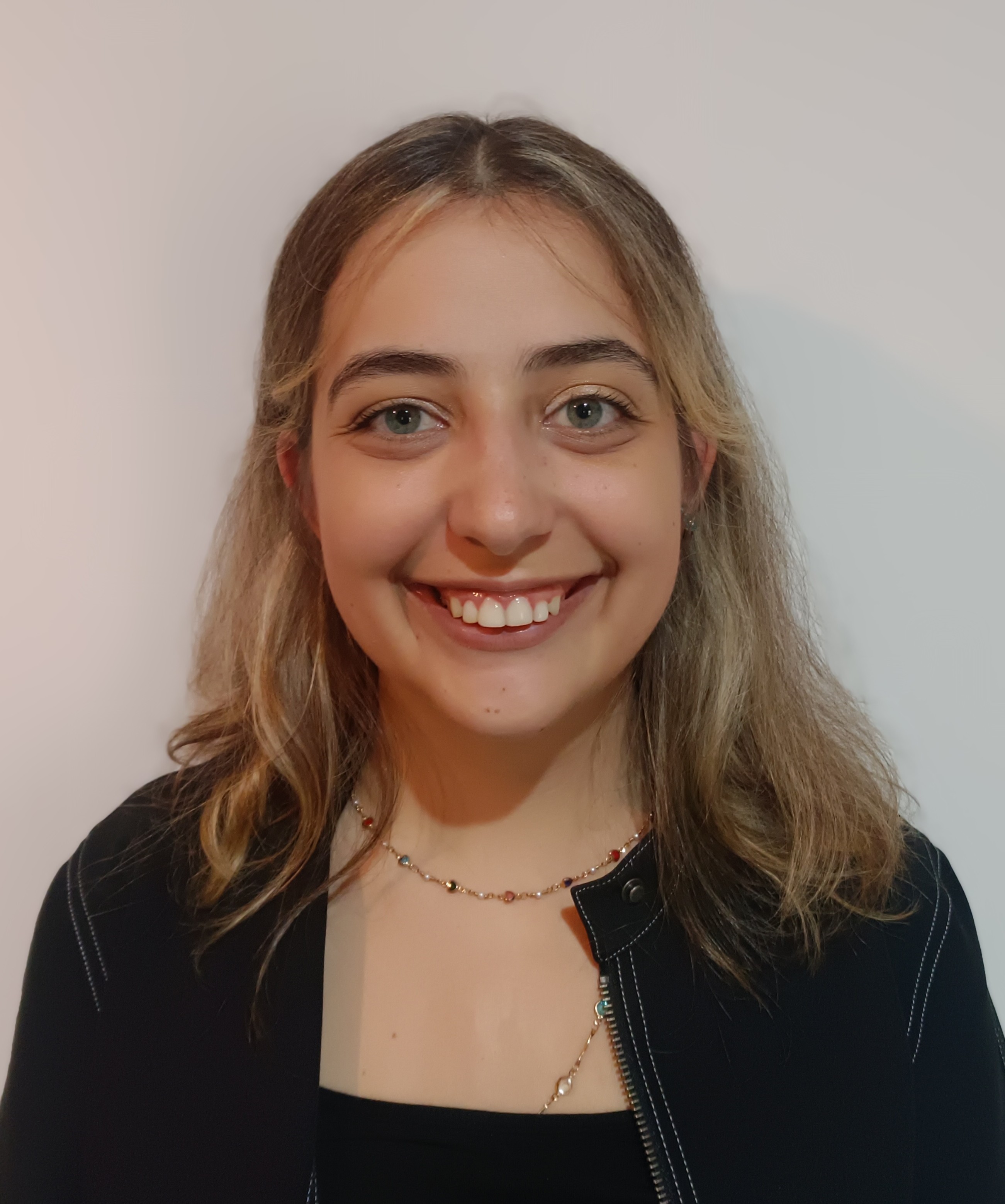}}]{VIVIANA CENTRITTO} received the B.S. degree in telecommunication engineering from Universidad Católica Andrés Bello in Caracas, Venezuela, in 2023, and the M.S. degree in advanced telecommunication technologies from Universitat Politècnica de Catalunya, Barcelona, Spain, in 2026. During her master's studies, she conducted research at the Nanonetworking Center in Catalunya (N3Cat) focusing on cryo-CMOS on-chip antenna design for quantum environments. Aditionally, she worked on the characterization of underwater wireless optical channels (UWOC) and reinforcement learning-based 3D beam adaptation for UWOC systems. She is currently a Ph.D. candidate in the Department of Digital Systems at the ICT School of the University of Piraeus, Greece, within the TeraWireless HORIZON-MSCA Industrial Doctoral Network. Her current research interests include statistical channel modeling of small scale fading channels, path resolvability, and the fundamental limit analysis of U-MIMO THz systems.
\end{IEEEbiography}

\begin{IEEEbiography}[{\includegraphics[width=1in,height=1.25in,clip,keepaspectratio]{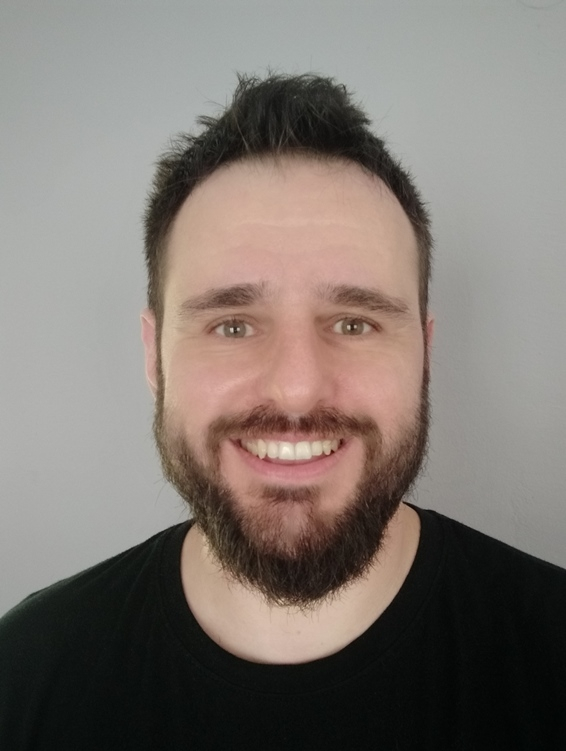}}]{SOTIRIS DROULIAS } received the diploma in electrical and computer engineering and the Ph.D. degree in nonlinear photonics from the National Technical University of Athens, Greece, in 2001 and 2007, respectively. From 2009 to 2012, he was an Adjunct Lecturer with the University of Patras, Greece, and from 2012 to 2020, he was a member with the Photonic-Phononic-and Meta-Materials Group, FORTH-IESL, Crete, Greece. He is currently a Research Associate with the Department of Digital Systems, ICT School, University of Piraeus, Greece. He is the author of more than 60 articles and four book chapters. He has worked on several EC funded projects. His research interests include electromagnetic modeling, metasurfaces, antennas, and propagation. In 2019, he received the Best Poster Award for his work on metasurface lasers in META 2019, Lisbon, Portugal, and in 2020, he was recognized as an Outstanding Reviewer from the Institute of Physics (IOP). He has received several talk invitations in prestigious conferences and he serves as a reviewer in international scientific journals.
\end{IEEEbiography}

\begin{IEEEbiography}[{\includegraphics[width=1in,height=1.25in,clip,keepaspectratio]{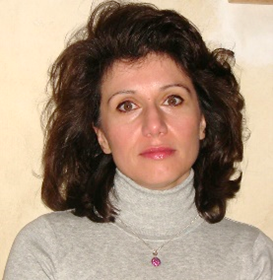}}]{ANGELIKI ALEXIOU } Angeliki Alexiou is a professor of Broadband Communications Systems at the department of Digital Systems, ICT School, University of Piraeus, Greece. She received the Diploma in Electrical and Computer Engineering from the National Technical University of Athens in 1994 and the PhD in Electrical Engineering from Imperial College of Science, Technology and Medicine, University of London in 2000. Since May 2009 she has been a faculty member at the Department of Digital Systems, where she conducts research and teaches courses on Broadband Communications and Advanced Wireless Technologies. Prior to this appointment she was with Bell Laboratories, Wireless Research, Lucent Technologies, (later Alcatel-Lucent, now NOKIA), in Swindon, UK, (January 1999-April 2009). Professor Alexiou is a co-recipient of Bell Labs President’s Gold Award in 2002 for contributions to Bell Labs Layered Space-Time (BLAST) project and the Central Bell Labs Teamwork Award in 2004 for role model teamwork and technical achievements in the IST FITNESS project. Professor Alexiou is the Chair of the Working Groups on Radio Communication Technologies and of on High Frequencies Radio Technologies of the Wireless World Research Forum and has been driving the WWRF contributions to ITU-R WP5D works on  IMT-2020  and IMT-2030. Her current research interests include radio interface for 6G systems, MIMO, THz wireless communication technologies, Reconfigurable Intelligent Surfaces, Joint Communications and Sensing, Near-Field Communications and Machine Learning for wireless systems. She is the technical manager of (SNS JU) INSTINCT project and the project coordinator of TeraWireless HORIZON-MSCA Industrial Doctoral Network.
\end{IEEEbiography}

\end{document}